\documentclass[aps,prb,twocolumn,superscriptaddress,floatfix]{revtex4-2}
\usepackage{amsmath}
\usepackage{graphics}
\usepackage{graphicx}
\usepackage{epsfig}
\usepackage{amssymb}
\usepackage{amsfonts}
\usepackage{color}
\usepackage{pifont}
\usepackage[normalem]{ulem}

\begin{document}

\title{Topological flat bands emerging at the inversion of stacking order in rhombohedral graphite}

\author{Ruben Weht}
\affiliation{Departamento F\'{\i}sica de la Materia Condensada, Gerencia de Investigaci\'on y Aplicaciones,
Comisi\'{o}n Nacional de Energ\'{\i}a At\'{o}mica (CNEA), \\
Avda General Paz y Constituyentes, 1650 San Mart\'{\i}n, Argentina}
\affiliation{Consejo Nacional de Investigaciones Cient\'{\i}ficas
y T\'ecnicas (CONICET), Buenos Aires, Argentina}
\affiliation{Instituto Sabato, Universidad Nacional de San Mart\'{\i}n - CNEA,
1650 San Mart\'{\i}n, Argentina}

\author{Armando A. Aligia}
\affiliation{Instituto de Nanociencia y Nanotecnolog\'{\i}a, CNEA-CONICET, GAIDI, Centro At\'{o}mico Bariloche and Instituto Balseiro, 8400 Bariloche, Argentina}

\author{Manuel N\'u\~nez-Regueiro}
\affiliation{Univ. Grenoble Alpes, CNRS, Grenoble INP, Institut N\'eel, 38000 Grenoble, France}

\date{\today }

\begin{abstract}
Motivated by the indications of high-$T_c$ superconductivity in natural graphite enriched in the rhombohedral phase, we study the band structure of several stacking configurations that combine two of the three graphite structures as well as modifications of the rhombohedral sequence (from ABCABC... to CBACBA...), using first-principles calculations. We focus in particular on the possible emergence of flat bands near the Fermi level. When the two different rhombohedral orderings are combined, flat bands of topological origin emerge at the interface between the two domains, near the $K$ and $K^\prime$ points of the Brillouin zone.
Mapping a simple tight-binding model of a rhombohedral slab along the direction perpendicular to the graphene layers onto a Su–Schrieffer–Heeger chain provides a transparent understanding of the underlying physics.
\end{abstract}

\maketitle

\section{Introduction}
\label{intro}

For more than two decades, there have been recurring claims of ferromagnetism and superconductivity above room temperature in various forms of graphite, including highly oriented pyrolytic graphite and natural single crystals with Bernal (ABAB…) and rhombohedral (ABCABC…) stackings~\cite{Kope00,Scheike12,Ariskina22}. 
While the origin of superconductivity in these compounds remains unclear, it has often been attributed to granular superconductivity at graphite interfaces~\cite{Pablo14}.

In particular, it has recently been shown that by using a magnetic field sorting method in suspensions of graphite grains, it is possible to concentrate the amount of superconducting particles, achieving percolation to zero resistance in electrical resistance measurements above room temperature.
Transmission electron microscope studies of the superconducting grains with $T_c \sim$ 650~K reveal a stripe-like contrast, suggesting a stacking of ordered two-dimensional domains with thickness of a few nanometers~\cite{manuel24}.
In contrast, magnetic separations performed on rhombohedral‑rich graphite yield powders that, according to X‑ray diffraction and Raman characterization, exhibit a large increase (up to about 40\% of the volume) in the rhombohedral phase~\cite{Champi} and a $T_c \sim$ 370~K.

Superconductivity has already been observed in twisted bilayer graphene~\cite{Cao18} and in rhombohedral few-layer graphene slabs~\cite{Zhou21,Zhang26}, although the reported transition temperatures $\mathrm{T_c}$ are quite small, of the order of a few Kelvin. An electron–phonon mechanism has been proposed for both systems~\cite{Wu19,Chou21}. In each case, the presence of flat bands appears to be the key ingredient for superconductivity. Moreover, the nontrivial topology of flat bands can enable relatively large values of $T_c$, scaling with the effective attractive interaction rather than exhibiting the exponential dependence characteristic of the conventional Bardeen–Cooper–Schrieffer framework~\cite{Peotta15,Torma22,Tian23}.

Using tight-binding formulations, it became clear that surface superconductivity develops at the surfaces of a slab of rhombohedral graphite, and that $T_c$ increases with the number of graphene layers~\cite{Kopnin11,Kopnin11b,Jiang26}. The localized surface states near the K and K$^\prime$ points of the two-dimensional $(k_x,k_y)$ Fermi surface and its topology can be explained by mapping the model in the third ($z$) direction to a Su–Schrieffer–Heeger (SSH) chain~\cite{Zhang24}. The geometrical properties of several arrangements of graphene layers have also been studied~\cite{Jiang26,liu25,Lv26}.

On the basis of these works, one may speculate that, in ordinary graphite, certain ordered stackings combining rhombohedral and possibly other domains (for example, through a twist at the interface) could give rise to flat bands similar to those associated with the surface states of finite rhombohedral slabs.
We have investigated the electronic structure of various graphite stacking configurations using first-principles density functional theory (DFT)~\cite{Hohe-Kon-1964,Kohn-1965}.
Similar studies have recently been performed using simplified tight-binding models~\cite{sun26}. However, a change between the two possible rhombohedral stackings (ABC… and CBA…) has not yet been addressed.
Analyzing the results using the effective SSH model, we conclude that this case is the only one in which robust
topological flat bands, similar to those of the surface of rhombohedral slabs are present.

The paper is organized as follows. In Sec.~\ref{met}, we describe the methodology used for the first-principles and tight-binding calculations. The results are presented in Sec.~\ref{res}. Section~\ref{sum} provides a summary and discussion.

\section{Methods of Calculation}
\label{met}

Throughout our work, we used the VASP code~\cite{VASP-1, VASP-2, VASP-3} to calculate energy bands, charges, density of states, etc., for graphite with various stacking configurations. It is  one of the most widely used Density Functional Theory (DFT) programs. It uses pseudopotentials to represent the interaction between electrons and ions, expressing both the wave functions and the potential as a combination of plane waves. The calculations were performed using the PBE exchange–correlation functional~\cite{PBE}, with van der Waals interactions treated using the Tkatchenko–Scheffler scheme~\cite{TS}.
Importantly, we took very fine meshes in reciprocal space to integrate charges and accurately determine the Fermi level.
Using this approach, we obtained interlayer separations in good agreement with the expected values for different graphene stackings: 3.35~$\mathrm{\AA}$ for AB (Bernal) graphite, 3.25~$\mathrm{\AA}$ for the ABC (rhombohedral) structure, and 3.53~$\mathrm{\AA}$ for AA stacking. 
The larger separation in the latter case is attributed to an enhanced repulsive interaction between carbon atoms aligned along the $z$ direction.

To interpret some of our results, we employ a simple tight-binding model with nearest-neighbor intralayer hopping $t$ (in the $xy$ plane) and interlayer hopping $t_z$ between atoms aligned along the perpendicular ($z$) direction in adjacent layers. Upon Fourier transforming in the
$xy$ plane, one obtains an effective model in terms of the states $\alpha _{i}(k)$ and $\beta _{i}(k)$, corresponding to the two inequivalent sites of layer $i$ with two-dimensional wave vector $k$.

\section{Results}
\label{res}

\begin{figure}[!b]
\centering
\includegraphics[width=8.cm]{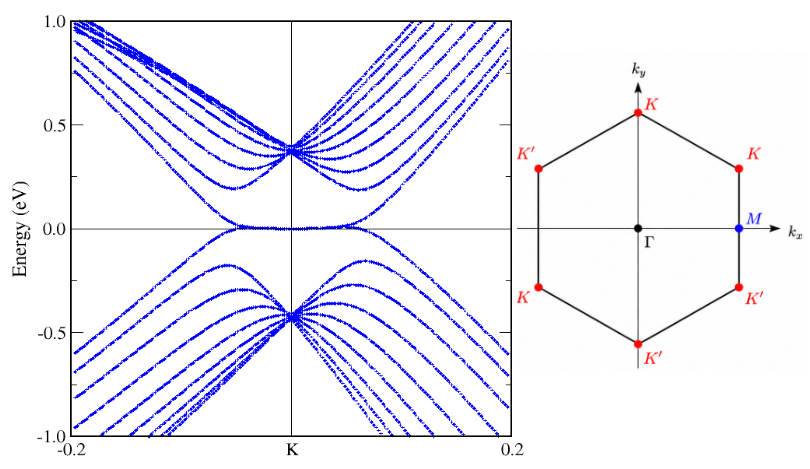}
\caption{Band structure along the $\mathrm{M}-\mathrm{K}-\Gamma$ lines in the reciprocal space for an isolated slab of 9 layers of graphene with the rhombohedral structure. We plot only an small region of the Brillouin zone around $\mathrm{K}$, units are $1/\mathrm{\AA}$.
On the right we include a diagram of the first Brillouin zone, highlighting the special points.}
\label{9r}
\end{figure}

To illustrate the kind of flat bands that we are looking for, we represent in Fig.~\ref{9r} the resulting bands for a slab with nine layers of graphene in a rhombohedral arrangement (ABC-ABC-ABC).
As we can see, there are two flat bands at the Fermi level around the $K$ point, in agreement with previous reports.
They correspond to states localized at both surfaces, with a small hybridization between them~\cite{Kopnin11,Kopnin11b,Jiang26}.

The result at the $K$ and $K^\prime$ can be readily understood from the tight-binding model. The states
$\alpha _{i}(k)$ and $\beta _{i}(k)$ do not hybridize at these high-symmetry points. Along the perpendicular direction, the interlayer hopping $t_z \sim$ 0.4 eV
hybridizes only the $\beta _{i}(K)$ states of layer $i$ with the
$\alpha _{i+1}(K)$ states of next one. For a slab of $N$ layers,
the states $\alpha _{1}(K)$ and $\beta _{N}(K)$ are decoupled,
yielding two states at the Fermi level (zero energy). The remaining states arise from  $N-1$ independent  $2 \times 2$ matrices of eigenvalues $\pm t_z$.
These states are clearly seen in the figure.

As it is evident from Fig. \ref{9r}, the states with energies $\pm t_z$ split out of the $K$ and $K^\prime$ points.
However, if the deviation of the wave vector $k$ from one of these
points is small, and if $N$ is large,
the zero-energy states persist due to the topology of the effective SSH model along the $z$ direction \cite{Zhang24} (see Supplemental Material \cite{supple}).

Do similar flat bands exist for a change in stacking in graphite layers?
We have found that there are no such flat bands at the interface
between AA ordering and either Bernal or rhombohedral ordering.

\begin{figure}[htb]
\centering
\includegraphics[width=7.cm]{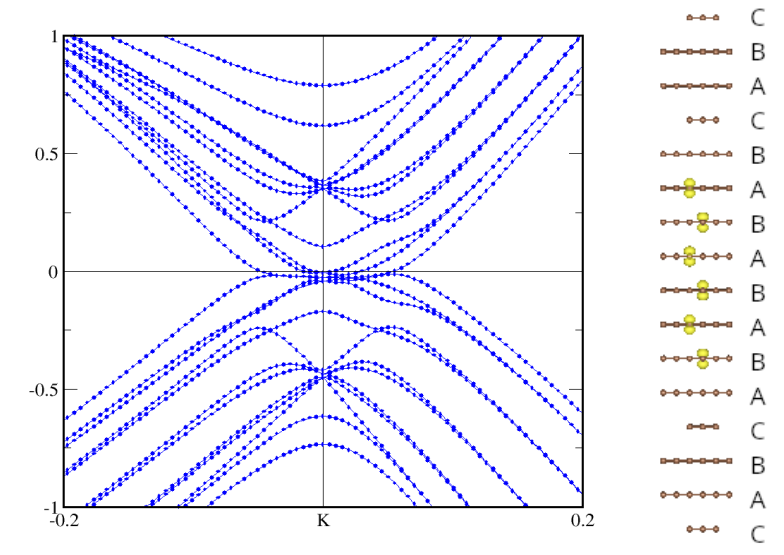}
\caption{Band structure along the $\mathrm{M}-\mathrm{K}-\Gamma$ lines and charge densities for energy levels around the Fermi level a compound of alternating six layers of Bernal type and six layers of rhombohedral type}
\label{abrhom2}
\end{figure}

In Fig.~\ref{abrhom2} we display the band structure for a stacking consisting  of six layers of both rhombohedral and Bernal structures,
and the charge distribution at low energies. Two flat bands
similar to those of Fig.~\ref{9r} are clearly visible near the Fermi level, in addition to additional structure.

Integrating the charge density in a small energy window of total width 0.1 meV around the Fermi level, we find that the states of the flat bands are spatially confined in the Bernal region. Flat bands in this system have also been obtained using tight-binding calculations in previous studies; however, those studies suggest charge localization at the interface between the two phases \cite{sun26}, which differs from our observations. This discrepancy could possibly arise from a change in the effective potential across the interface, an effect that may not be fully captured within the tight-binding framework.
The resulting charge distribution is expected from the mapping of a finite rhombohedral slab onto the SSH model \cite{Zhang24,supple}. When the slab’s edge states hybridize with Bernal states near the $K$ point (where the latter can be described as a tight-binding chain with uniform hopping $t_z$) the rhombohedral edge states broaden in energy and become embedded within the Bernal structure.

\begin{figure}[htb]
\centering
\includegraphics[width=7.cm]{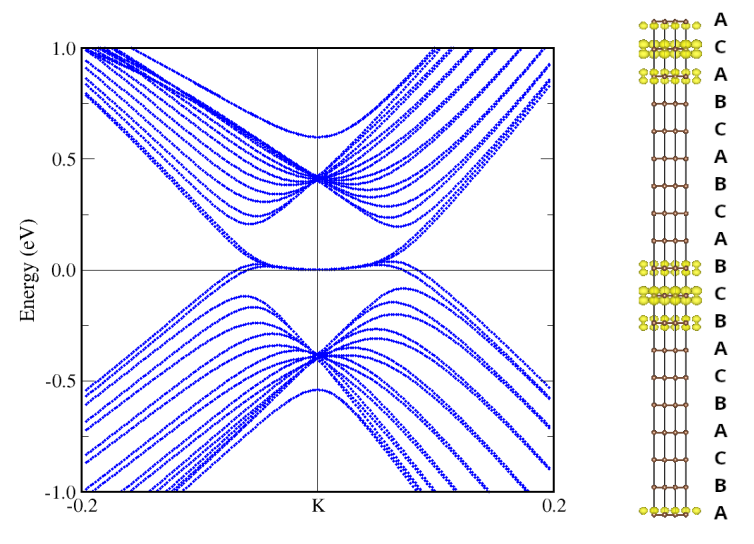}
\caption{Band structure along the $\mathrm{M}-\mathrm{K}-\Gamma$ lines and charge densities at the $K$ point,  for a compound of
alternating nine layers of rhombohedral type ABC... and nine layers with the alternative stacking BAC.}
\label{abcba}
\end{figure}

Finally, we consider the central result of this work in which slabs with opposite rhombohedral stackings are joined. The corresponding band structure is shown in Fig.~\ref{abcba}, where four flat bands near the Fermi level are clearly visible around the $K$ point.
In contrast to the previous case, the charge density
of the bands nearest to the Fermi level (four in our case) is now concentrated at the layer (denoted $N$) where the rhombohedral stacking reverses, as well as in the two adjacent layers. Away from the $K$ point, the charge density associated with the four states near the Fermi level becomes more delocalized within the sample, and for $|k-K| \sim 0.05$ it is distributed across all layers.

\begin{figure}[t!b]
	\centering
	\includegraphics[width=4.cm]{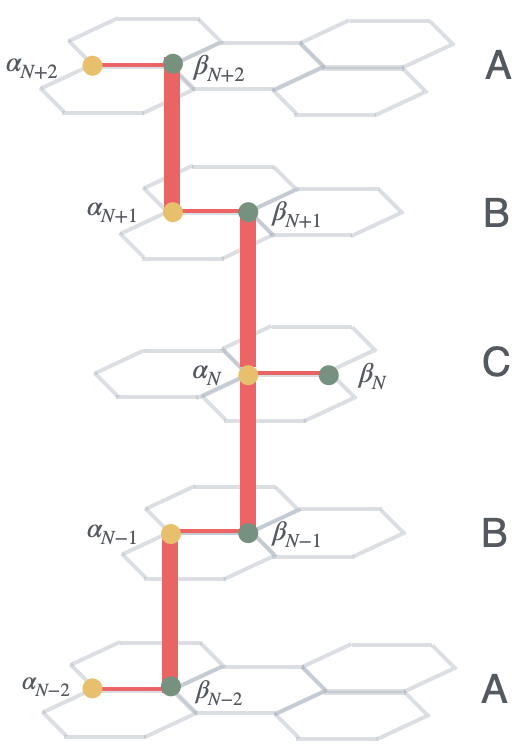}
	\caption{Schematic illustration of the atomic structure of rhombohedral
	graphite near the change of stacking and the effective hoppings between
	Bloch states.}
	\label{solit}
\end{figure}

As in the case of a rhombohedral slab, these results can be understood using a simple tight-binding model with nearest-neighbor in-plane hopping $t$ and interlayer hopping $t_z$ between atoms aligned in adjacent layers.
Assume that at a layer $i=N$, corresponding to C, the
rhombohedral ordering of the chains is inverted,
so that near $i=N$ the ordering of the graphene sheets is ... ABCBA ... The effective model for the system
in the $z$ direction takes the form of two SSH chains,
one for $i \leq N-1$ and another for $i \geq N+1$
hybridized with the states $\alpha _{N}(k)$ and $\beta _{N}(k)$ as represented in Fig. \ref{solit}.
Note that for an infinite system, there is a mirror symmetry $M$ around the layer C. In other words, the
system is invariant under the substitution $i\leftrightarrow 2N-i$.

We discuss first the physics at the $K$ point for which
the hopping between $\alpha _{i}(k)$ and $\beta _{i}(k)$, denoted by $t_{k}$, vanishes, and
the only relevant hopping is $t_{z}$. In this case,
$\beta_{N-1}(K)$ hybridizes only with $\alpha _{N}(K)$, and as expected by the
symmetry $M$,  $\beta _{N+1}(K)$ also hybridizes only with $\alpha _{N}(K)$.
The antisymmetric contribution $\beta _{N-1}(K)-\beta _{N+1}(K)$ does not
hybridize with any state and remains at zero energy.
In addition, the state $\beta _{N}(K)$ does not
hybridize and gives a second state at zero energy.
Naturally, the same happens at the $K^\prime$ point.

For small $q=k-K$, one has to linear order in $q$ \cite{supple}
\begin{equation}
t_{k}=(3ta/2)(-q_{x}+iq_{y}).  \label{tq}
\end{equation}
The phase can be absorbed by a gauge transformation.
One knows that the SSH chain is topological as long as
$|t_{k}|<t$, or using Eq. (\ref{tq})
\begin{equation}
|qa|<\frac{\sqrt{2}}{3}|\frac{t_{z}}{t}|.  \label{qa}
\end{equation}

Therefore, as long as Eq.~(\ref{qa}) is satisfied, topological zero-energy edge states localized near
$\beta _{N-1}(K)$ and $\beta _{N+1}(K)$ exist.
An analytic expression for these states is given in Section III of Ref. \cite{Ali25}. The antisymmetric combination of these states remains unaffected by hybridization with $\alpha _{N}(K)$.

In addition, the state $\beta _{N}(k)$ hybridizes only in the plane and with
state $\alpha _{N}(k)$. Including the state (even under $M$) $\beta _{e}=(\beta
_{N-1}(k)+\beta _{N+1}(k))/\sqrt{2}$, the Hamiltonian matrix in the sector even under $M$ for $k \rightarrow K$ takes the form

\begin{equation}
H=\left(
\begin{array}{ccc}
0 & -t_{k} & 0 \\
-\bar{t}_{k} & 0 & \sqrt{2}t_{z} \\
0 & \sqrt{2}t_{z} & 0
\end{array}
\right) .  \label{h3}
\end{equation}

For $k\neq K$, the state $\beta _{e}$ should be replaced by the
corresponding symmetric combination of the SSH end states (as long as Eq. (\ref{qa}) is satisfied) and the magnitude of the matrix elements is
reduced, but the form of the matrix is retained.
There is an eigenstate of
zero energy proportional to $\sqrt{2}t_{z}\beta _{N}(k)+t_{k}\beta _{e}$.

\begin{figure}[t!b]
	\centering
	\includegraphics[width=7.cm]{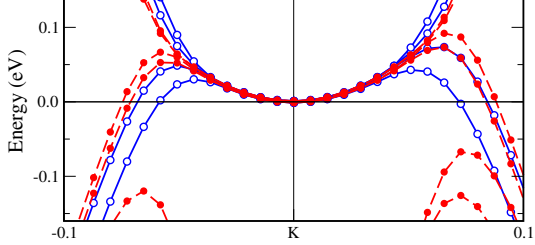}
	\caption{Zoom of the band structures along the $\mathrm{M}-\mathrm{K}-\Gamma$ lines around the Fermi level for $n$ABC-$n$BAC slabs with $n=3$ (solid lines, empty circles) and $n=5$ (dashed lines, filled circles).}
	\label{bands-3-5}
\end{figure}

The above discussion assumes a single defect in an otherwise very large system. If multiple boundaries or stacking changes are present, the corresponding SSH chains become finite, and the end states hybridize (see, e.g., Sec.~III of Ref.~\cite{Ali25}). As a result, the range of $q$ values for which low-energy localized states exist is reduced. To illustrate this fact, we display
in Fig. \ref{bands-3-5} how the low-energy bands change when the distance between consecutive interfaces
changes from 9 to 15 layers. In the latter case, the bands have a smaller splitting and deviate from a nearly flat behavior at larger values of $|k-K|$.

In summary, near the $K$ point, two states protected by topology, one even
and the other odd under mirror symmetry, are localized near zero energy and
near the layer of the change of stacking.
Since the unit cell contains two layers where the stacking changes, four low-energy bands are expected, in agreement with Fig.~\ref{abcba}. In addition,
from the matrix Eq. (\ref{h3}) for $k=K$ one obtains two eigenvalues at
$\pm \sqrt{2}t_{z}$ while most of the other eigenvalues at this point lie at
$\pm t_{z}$. This fact agree with the four split bands (two per interface) observed in Fig.~\ref{abcba} at $k=K$.
The distribution of charges in the figure is also fully consistent with the above discussion.

\section{Summary and discussion}
\label{sum}

We have investigated several interfaces between different stackings of graphene layers as candidate structures in graphite. Previous studies of rhombohedral graphene slabs have shown the presence of topologically protected flat bands at their surfaces. In particular, for sufficiently strong effective electronic attraction, the superconducting critical temperature $T_c$ increases approximately linearly with the slab thickness,
up to about 20 layers \cite{Kopnin11b}.

Based on our results, we expect similar physics to arise at interfaces between the two possible rhombohedral stackings. Our first-principles calculations indeed reveal flat bands localized at such interfaces. Furthermore, using a tight-binding model that captures the essential physics, we find that near the $K$ and $K^\prime$ points of the two-dimensional Brillouin zone these states can be understood in terms of the end states of an SSH model in its topological phase, exhibiting both similarities and differences with the surface states of rhombohedral slabs.

As the distance between interfaces decreases, the states associated with different interfaces hybridize, leading to the opening of a small gap between the flat bands. This behavior can also be understood in terms of the SSH model for a finite chain \cite{Ali25}. The hybridization further increases as the two-dimensional wave vector $k$ moves away from the $K$ or $K^\prime$ points. For sufficiently large deviations [when Eq.~(\ref{qa}) is no longer satisfied] the effective SSH model undergoes a transition to a topologically trivial phase, and the localized low-energy states disappear.

In addition to the expected increase in $T_c$ with the distance between
interfaces, we we anticipate a strong dependence on pressure. Increasing pressure reduces the interlayer distance $d$ between adjacent layers, and the magnitude of the hopping $t_z$ between $2p_z$ orbitals is expected to increase as
$1/d^3$ \cite{harrison}. In the strong-coupling limit, $T_c$ is
proportional to the area of the region of the flat band \cite{Kopnin11b},
which according to Eq.~(\ref{qa}), scales as $t^2_z$. Therefore, one expects
$T_c \sim 1/d^6$.

We also obtain flat bands for the interface between Bernal and rhombohedral stackings. However, the low-energy states are not localized at the interface according to our results.

Concerning experiments, artificially fabricated stackings of twisted or rhombohedral graphene layers have so far exhibited superconductivity only at low temperatures. Although these measurements have been reproduced by several groups, the success rate remains low, as the samples are difficult to fabricate due to factors such as structural disorder (e.g., buckling or irregular stacking), large interlayer separations, and impurities. In addition, superconductivity is highly sensitive to the level of doping.

In contrast, such defective superconducting structures may occur naturally and randomly in graphite \cite{Kope00,Scheike12,Ariskina22}. When the appropriate stacking order and doping are present, they can be selectively concentrated using magnetic-field-based separation techniques that exploit their large diamagnetism \cite{manuel24}. These selected regions exhibit ordered stackings, at least in samples with the highest reported
$T_c$ values.

The presence of topological flat bands at interfaces between the two rhombohedral orderings is expected to promote superconductivity, as in the case of the surfaces of rhombohedral slabs, and possibly with higher $T_c$.
Outer surfaces are often highly irregular, with Bernal or even AA-type defects that induce scattering between different wave vectors $k$, thereby disrupting the flat bands. In contrast, interfaces are typically sharp and well defined.

Such an interface can be engineered by cutting a rhombohedral graphite stack of approximately $\sim 20$ layers or more and placing one half on top of the other after a 180° rotation. There are three possible ways of cutting the stack, depending on whether the terminating layer is of type A, B, or C, and three corresponding ways of rejoining the two pieces, depending on their relative displacement. Disregarding the higher-energy AA, BB, and CC contacts, the resulting interface configuration is equivalent to the one studied here, up to a cyclic permutation of the ABC stacking sequence.

\section*{Acknowledgments}
R.W. sincerely thanks Joaquín Torres and Bruno Bruzzo for their computational assistance, as well as the partial support of SICyT and the management team for supercomputing time on Clementina XXI (grant No~PAD-148-2025).

\bibliography{ref}

\end{document}